# MAGNETOSPHERIC DYNAMICS AND CHAOS THEORY[1]

G.P. PAVLOS

Democritus University of Thrace

*Abstract*

The results of this study were announced and published in Greek in the Fifth Panhellenic Conference Proceedings of the Hellenic Physical Society. It is the sequel of a previous study (Pavlos, 1988), in which it was introduced the hypothesis of magnetospheric chaos for the interpretation of magnetic substorms.

In this study it is described the possibility of tracing magnetospheric chaos through Grassberger and Procassia method for the estimation of correlation dimension. In addition, it is proposed, the estimation of chaoticity through the computation of Lyapunov exponents. This study and its previous one constitute the first studies ever concerning the hypothesis of magnetospheric chaos for the interpretation and understanding the magnetospheric substorms. A series of publications of G.P.Pavlos followed the initial two studies in scientific journals and conference proceedings (www.gpavlos.gr). The publication of this study in English version has a historical importance and interest regarding the history of evolution of the concept of magnetospheric chaos. For an extended discussion concerning magnetospheric chaos, see, Pavlos 2012 ArXiv.

## 1. Magnetospheric Dynamics

*The Earth's Magnetosphere constitutes one of the most significant plasma laboratories, in which in situ analysis and studies of various solar and stellar or galactic plasma processes can be made (Pavlos, 1988). The basic problem of*

---

[1] This study in its original form (section 1-4) was presented and published in the proceedings 5th HELLENIC PHYSICS CONFERENCE ATHENS, 1989

magnetospheric dynamics, which till now isn't fully theoretically comprehended, is the bursting release of magnetic energy during magnetospheric substorms, which follow the southern turn ($B_z < 0$) of the interplanetary magnetic field (MacPherron, 1987; Pavlos, 1988). In this case, the growth rate of the magnetic field accumulation, $E_{magn} \sim B^2/2\mu_0$, is increased inside the Earth's magnetosphere, resulting the lengthening-thinning of Earth's magnetotail and driving the whole magnetopsheric system into an unstable state.

The magnetopsheric substorm includes a series of various phenomena which go with the development of plasma instability, with main result the cutoff and the ejection of great masses of magnetospheric plasma towards the direction of the solar wind flow (Fig.1). Characteristic phenomena which have been repeatedly observed in Earth's magnetosphere during magnetospheric substorms are discriminated into: a) Polar – Ionospheric phenomena such as acceleration and superposition of energetic particles in the Polar Regions, luminous phenomena (auroral borealis), enforcement of ionospheric currents, influx of plasma and energetic particles in the equatorial ionosphere, efflux of ionospheric ions from magnetotail, radiation, iono-cyclotron waves and b) Distant magnetospheric phenomena: intense topological variations of the magnetospheric magnetic field, development of powerful local electric fields which accelerate charged particles till 1-2 MeV (bursts of energetic particles), generation of intense plasma flows towards or opposite to Earth with a velocity of ~ $10^9$ km/sec (Pavlos PhD Thesis, Democritus University of Thrace, 1983). The theoretical models that were proposed in the past in order to interpret physically the above phenomenology of the magnetospheric substorms, can be discriminated into two groups: 1) Loading-Unloading process, based on Hones model or the X-neutral line model (Hones, 1979; Pavlos & Sarris, 1989). 2) Driven process which is represented by models of boundary layers (Rostoker, 1987).

In this study, we try to study the behavior of the magnetospheric system and its dynamics from chaos theory's angle. This attempt can be justified from the fact that the coupling of Earth's magnetosphere with the solar wind as well as the ionosphere, leads to a nonlinear magneto-hydrodynamic system with a strong holistic character.

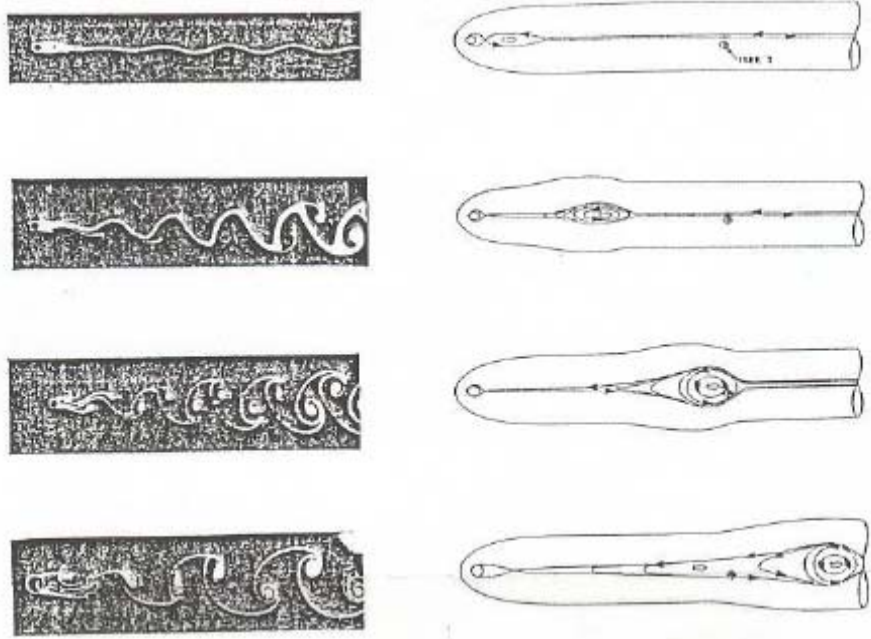

***Fig.1.*** *Analogy between the evolution of the turbulent flow in a fluid and the flow of magnetospheric plasma along the magnetotail during magnetopsheric substorms.*

## 2. Chaotic Dynamics

*The microscopic description of magnetosphere would require infinite degrees of freedom (description of electromagnetic field and charged plasma particles). On the contrary, the macroscopic description can be achieved with finite degrees of freedom. Supposing $x^1(t), x^2(t), \ldots, x^N(t)$, where $x^i(t), i = 1, \ldots, N$, the necessary physical variables of the system. The dynamics of a dissipative system, such as the magnetospheric plasma, we assume that can be given through a system of nonlinear differential equations of the form:*

$$\frac{d\mathbf{x}^i}{dt} = \mathbf{F}(\mathbf{x}^i(t), \mu), i = 1, \ldots, N \qquad (1.1)$$

*(Eckman and Ruelle, 1985), where the parameter $\mu$ describes the dependence of the magnetosphere from the solar wind or/and the lower ionosphere – atmosphere. The variation of the parameter $\mu$ which corresponds to the enforcement of the external coupling leads the system to the emergence of intrinsic oscillations till the state of turbulence, through the bifurcation process. For the*

*nonlinear dissipative systems, this variation of the parameter $\mu$ usually goes with the emergence of initially a limit cycle attractor, afterwards a torus attractor and finally a strange or chaotic attractor. Before the appearance of the chaotic attractor the system has a deterministic behavior, while in the state of the strange attractor there is stochastic behavior. Correspondingly, the power spectrum of one of the measured variables $x^i(t) = u(t)$ it is linear before the shift to the strange attractor state, whereas in the chaotic attractor state becomes continuous. According to Wiener-Khinchin theorem, the power spectrum $S(\omega)$ of the quantity $u(t)$, which is given from the square of the Fourier transformation of the $u(t)$, equals to the Fourier transformation of the autocorrelation function of $u(t)$. Thus:*

$$S(\omega) = (\mathbf{const})\lim_{\tau \to \infty}\frac{1}{\tau}\left|\int_0^\tau dt e^{i\omega \tau}u(t)|u(t)|^2\right. =$$
$$= (\mathbf{const})\int_{-\infty}^{+\infty} dt e^{i\omega \tau} \lim_{\tau \to \infty}\frac{1}{\tau}\int_0^\tau d\tau u(\tau)u(\tau + t) \qquad (1.2)$$

*The stochastic behavior of the system in the chaotic attractor state, is due to the sensitivity to initial conditions, which correspond to the existence of at least one positive Lyapunov exponent of the flow matrix $D\mathbf{F}(\mathbf{x}(t))$ where $\mathbf{x}(t) = (x^1(t), \ldots, x^N(t))$. This means that the distance $\delta\mathbf{x}(t)$ of two nearby initial states (no matter how close they are) increases exponentially, according to the equation:*

$$\frac{d}{dt}\delta\mathbf{x}(t) \cong D\mathbf{F}(\mathbf{x})\delta(\mathbf{x}) \qquad (1.3)$$

*The chaotic attractor is a subspace of the phase space $R^N$ and with a noninteger dimension, called as fractal dimension. In Chaos theory we come across into 3 definitions of chaotic attractor's dimension: the capacity dimension, the information dimension and the correlation dimension. The last definition plays a crucial role in the experimental tracing of a chaotic attractor and is defined by the relation:*

$$D_c = \lim \frac{\ln C(\varepsilon)}{\ln(\varepsilon)} \qquad (1.4)$$

*where*

$$C(\varepsilon) = \lim_{N \to \infty}\frac{1}{N^2}\left\{\mathbf{x}_i, \mathbf{x}_j : \|\mathbf{x}_i - \mathbf{x}_j\| < \varepsilon\right\} \qquad (1.5)$$

## 3. The Relation of Attractors Theory with the Magnetospheric Dynamics

*In this paragraph we will show two ways in which we can interpret-describe the magnetospheric dynamics based on Chaos theory and on the development of chaotic attractors. One basic parameter of the magnetosphere is the rate of solar plasma energy input E(t) defined by the relation:*

$$E(t) = VB^2 \sin^4(\theta/2) l_0 \qquad (1.6)$$

*where V is the velocity of solar wind, $l_0 \cong 7R_e$ ($R_e$ = 1 Earth radius), B = the intensity of the magnetic field and θ the angle that is formed by the component $B_{yz}$ and the axis Z in the solar elliptic system ($x_{yz}$).*

*For values $E(t) < 10^{18} erg/\sec$, the magnetospheric system remains in its equilibrium state, while for $\delta(\tau) > 10^{18} erg/\sec$ there are intensive intrinsic oscillations (modes) which lead to the emergence of magnetospheric substorms. In this case the existence of a chaotic attractor can be traced with the Grassberger & Procassia method (1983). Based on this method we estimate the number $C(\varepsilon, N)$ of pairs of states in the topological equivalent to the initial phase space, $R^N$. This space $R^N$ can be constructed from the experimental time series corresponding to one of the macroscopic variables, for example $\{u(t), u(t+\tau), \ldots, u(t+N\tau)\}$, where the τ is chosen properly (Tsonis & Elsner, 1989; Parker and Chua, 1987). The existence of chaotic attractor as well as the correlation dimension $D_c$ can be derived from the relation $C(\varepsilon, N) \cong E^{D_c}$, for a finite number of N $D_c$ is stabilized for any increase of the dimension N of the embedding dimension of the experimental time series.*

*Another way according to which, the possible existence of chaotic attractor in the magnetospheric substorms can be examined, derives from the possibility of chaotic behavior of the equivalent electric circuit which models the electric currents of magnetospheric system, as given for example by Bostrom (1974). The chaotic behavior in this case, is related to the nonlinear behavior of equivalent resistors, either of the Ionosphere or the plasma sheet.*

*4. Conclusions*

*In this study we showed two possible ways, from which it is possible to relate the dynamics of magnetospheric substorms with Chaos theory. The ground and satellite measurements can reveal the truth concerning these hypotheses.*

**5. Discussion and new results**

This study followed after the introduction of the Chaos hypothesis for the magnetospheric dynamics by (Pavlos, 1988). During the period 1990-2000 the chaos hypothesis was founded in a series by (Baker, 1990), (Vassiliadis, 1991), (Chang, 1992), (Pavlos, 1991;1992). Pararelly strong criticism was developed against magnetospheric chaos by many scientists while the self-organized criticality (SOC) was proposed antithetically to chaos. (Pavlos et al 1994; 1999; 2000; 2001) and (Athanasiou and Pavlos, 2000; 2001) developed an extended non-linear data analyze algorithm by using the Singular decomposition analysis (SVD) method and the Theiler surrogate data method and concluded as an inevitable result the existence of a low dimensional chaotic component in the magnetospheric data on Earth (AE index) or in situ by space craft observations. Moreover Pavlos and Athanasiou tested the magnetospheric chaos hypothesis after the comparison of experimental data analysis and simulated data analysis obtained by using stochastic models. Recently Pavlos et al (20011; 2012) by analyzing spacecraft data existence of magnetospheric phase transition process during substrom periods, the magnetospheric phase transition process includes a SOC state during calum periods and low dimensional chaos state during substrom periods. Also during substorm periods we observe non0Gaussian statistics of the magnetospheric data simultaneously with the development of Tsallis non-extensive states of the magnetospheric system (Pavlos, 2011). Moreover Tsallis q-statistics anomalous diffusion and fractal dynamics are the inevitable underlying characters of the low dimensional magnetospheric chaos as it was shown by Pavlos (2012a; 2012b, 2012c). Moreover the magnetospheric chaos hypothesis has been extended and has been included in the non-equilibrium strange dynamics of the magnetospheric system. The non-equilibrium strange dynamics can be manifested: a) by low dimensional chaotic process (fig.2) b) by Tsallis non-extensive statistics and non-equilibrium transition process (fig.3, 4). These new theoretical and experimental results can be used for the theoretical understanding of magnetospheric processes as earthword-tailword bulk plasma flows (fig.5, 6) and magnetospheric bursts of energetic particles (fig.7) which have been observed many years before. Finally as we indicate in recent studies (Pavlos, 2012) the old mechanism of the magnetic reconnection process must be corrected by a new mechanism of fractal dissipation and fractal acceleration.

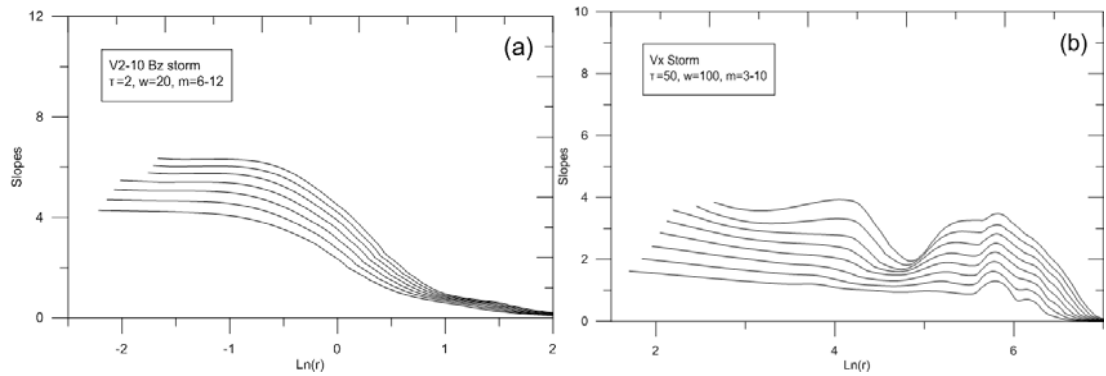

**Fig.2. (a)** presents the slopes of the correlation integrals estimated for the Bz time series and **(b)** for the Vx time series during magnetospheric periods. The low value saturation of the slopes reveals the development of a magnetospheric low dimensional strange attractor during the substrorm period.

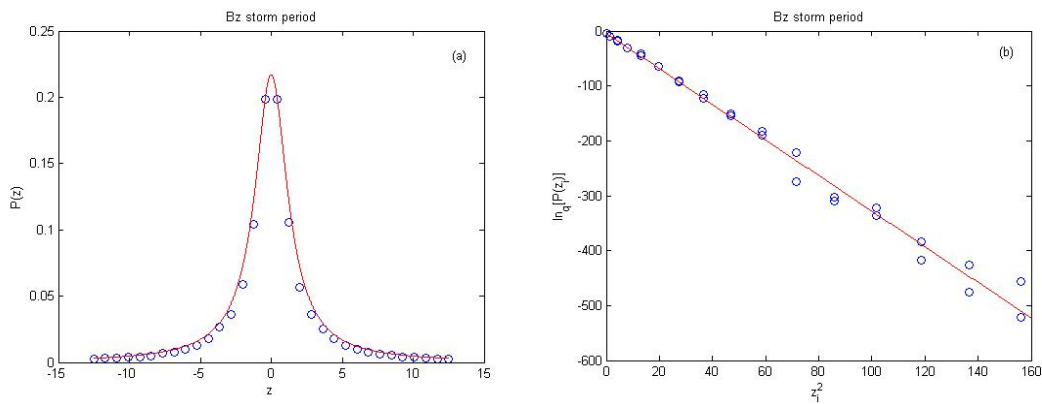

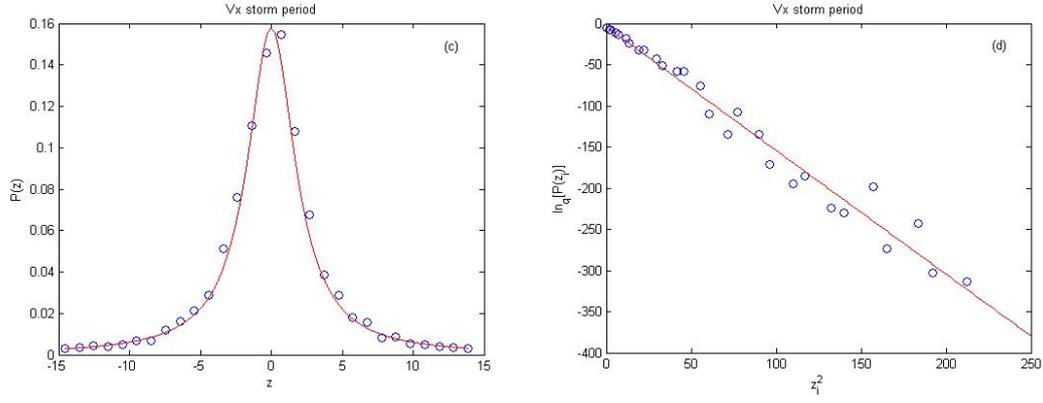

***Fig.3. (a)*** *PDF P($z_i$) vs. $z_i$ q Gaussian function that fits P($z_i$) for the Bz storm period **(b)** Linear Correlation between $ln_q P(z_i)$ and $(z_i)^2$ where q = 2.05 ± 0.04 for the Bz storm period **(c)** PDF P($z_i$) vs. $z_i$ q Gaussian function that fits P($z_i$) for the Vx storm period **(d)** Linear Correlation between $ln_q P(z_i)$ and $(z_i)^2$ where q = 1.98 ± 0.06 for the Vx storm period.*

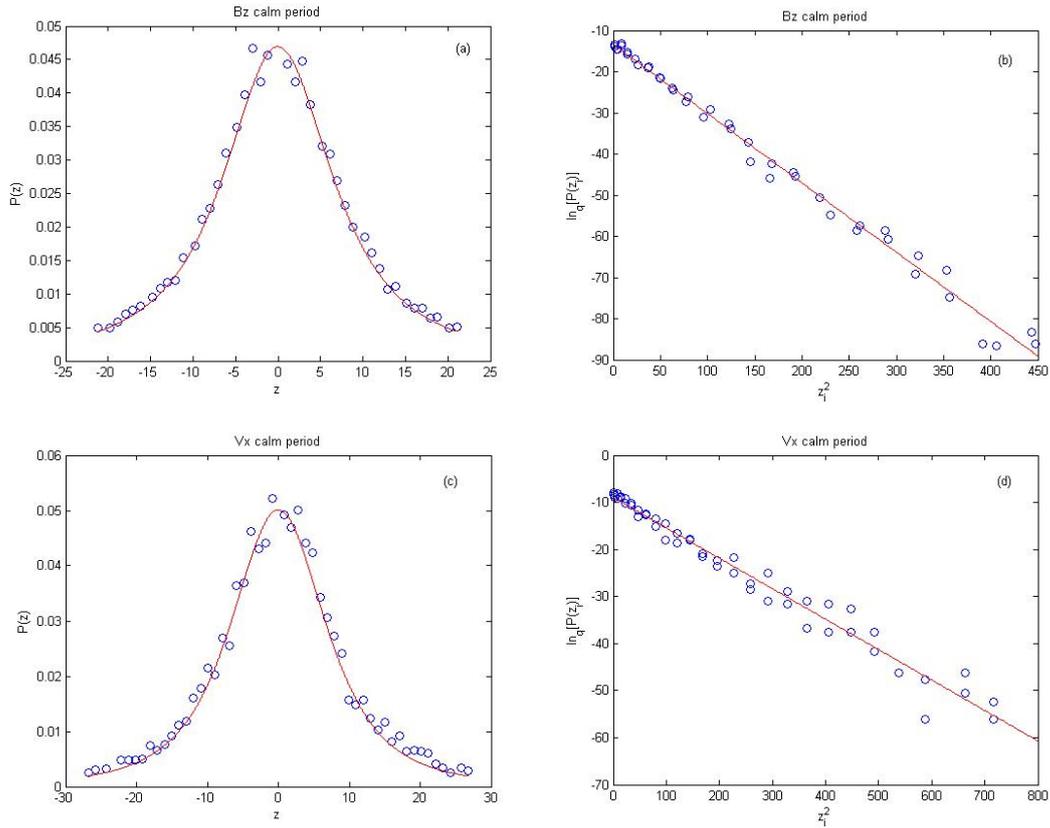

***Fig.4. (a)*** *PDF P($z_i$) vs. $z_i$ q Gaussian function that fits P($z_i$) for the Bz calm period **(b)** Linear Correlation between $ln_q P(z_i)$ and $(z_i)^2$ where q = 1.8 ± 0.06 for the Bz calm period **(c)** PDF P($z_i$) vs. $z_i$ q Gaussian function that fits P($z_i$) for the Vx calm period **(d)** Linear Correlation between $ln_q P(z_i)$ and $(z_i)^2$ where q = 1.59 ± 0.07 for the Vx calm period.*

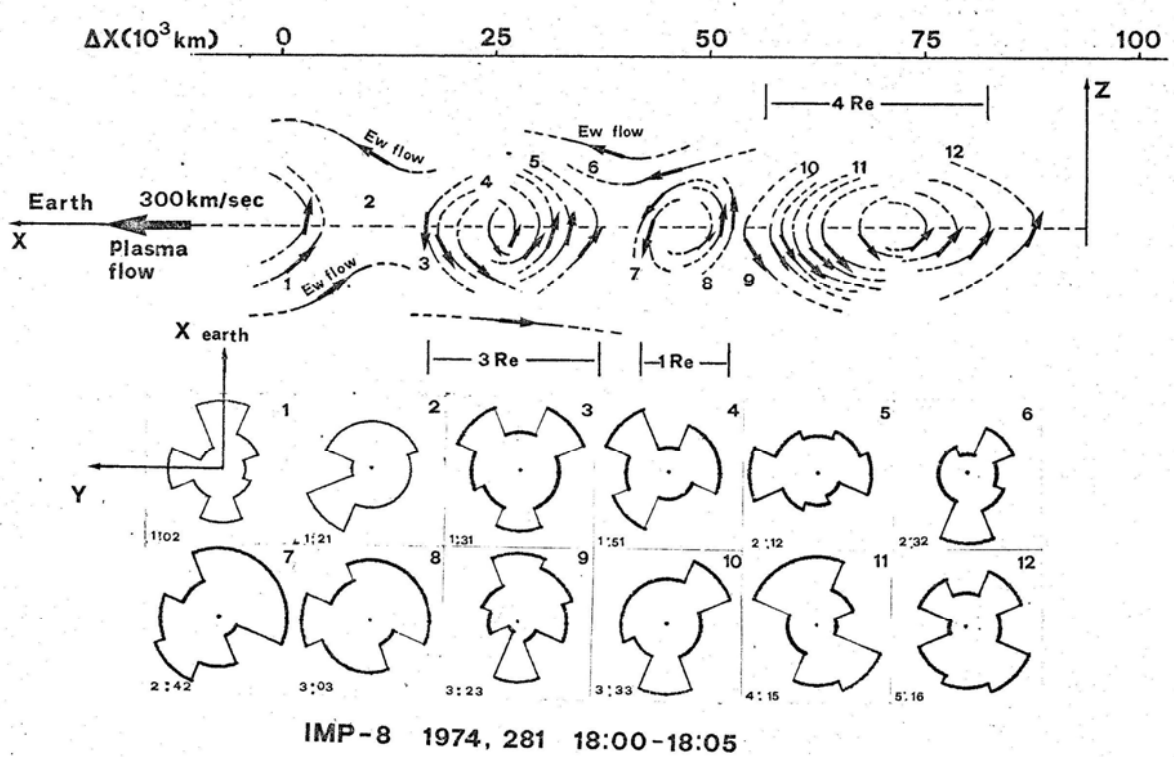

**Fig.5** In this figure we present an event of earthword bulk plasma flow in the earth plasma sheet. The angular distribution of the energetic electrons shown (bottom panel) in the same figure reveal closed magnetic lines (dublu pick profile) or green magnetic lines (anisotropic angular distribution profile). The topology of the magnetotail plasma sheet magnetic field is presented also in the (top panel). The representation of the magnetic field topology simmultanesly with the bulk plasma flow reveal local reconnection process which producesclosed magnetic lines structures and energetic particles. This phenomenon is also in agreement with the new theoretical concepts of fractal acceleration and fractal dissipation as the magnetic field and energetic particle distributions are clearly non-Gaussian and belongs to q-statistics profile with q>1.

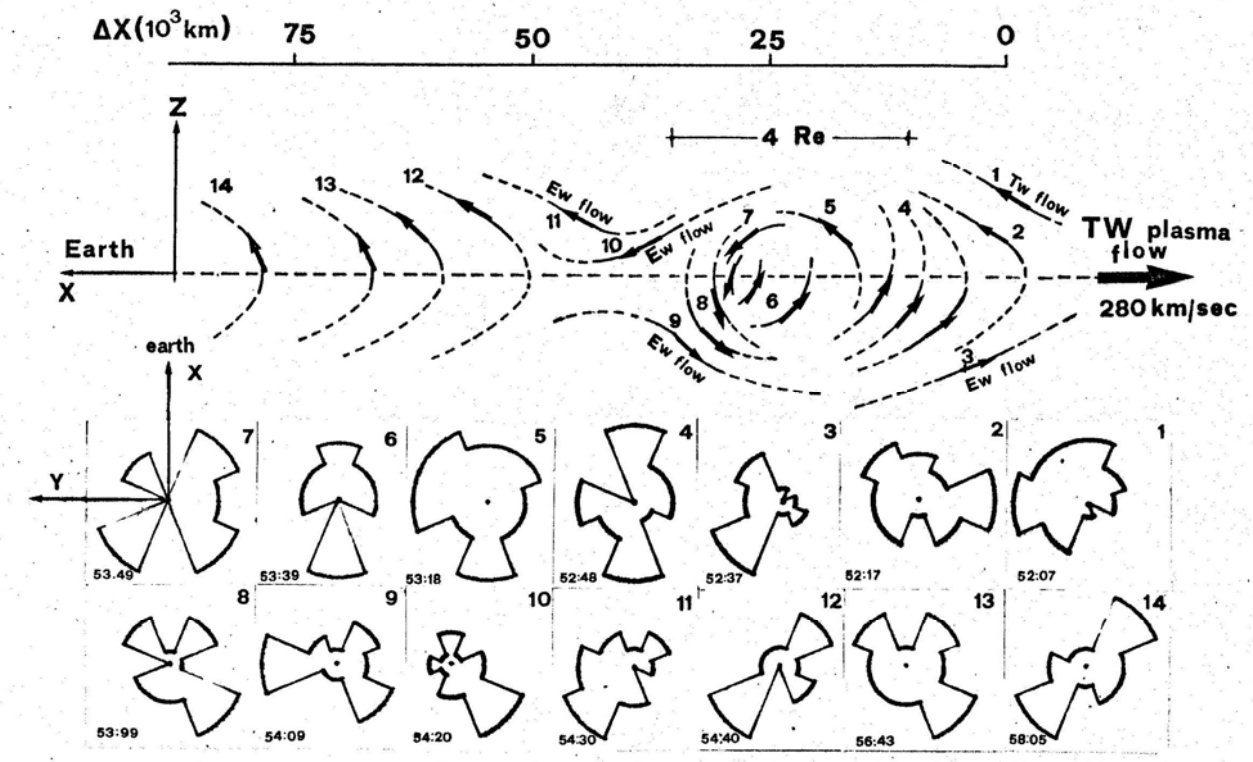

**Fig.6** In this figure we present an event of tailword bulk plasma flow in the earth plasma sheet. The angular distribution of the energetic electrons shown (bottom panel) in the same figure reveal closed magnetic lines (dublu pick profile) or green magnetic lines (anisotropic angular distribution profile). The topology of the magnetotail plasma sheet magnetic field is presented also in the (top panel). The representation of the magnetic field topology simmultanesly with the bulk plasma flow reveal local reconnection process which producesclosed magnetic lines structures and energetic particles. This phenomenon is also in agreement with the new theoretical concepts of fractal acceleration and fractal dissipation as the magnetic field and energetic particle distributions are clearly non-Gaussian and belongs to q-statistics profile with q>1.

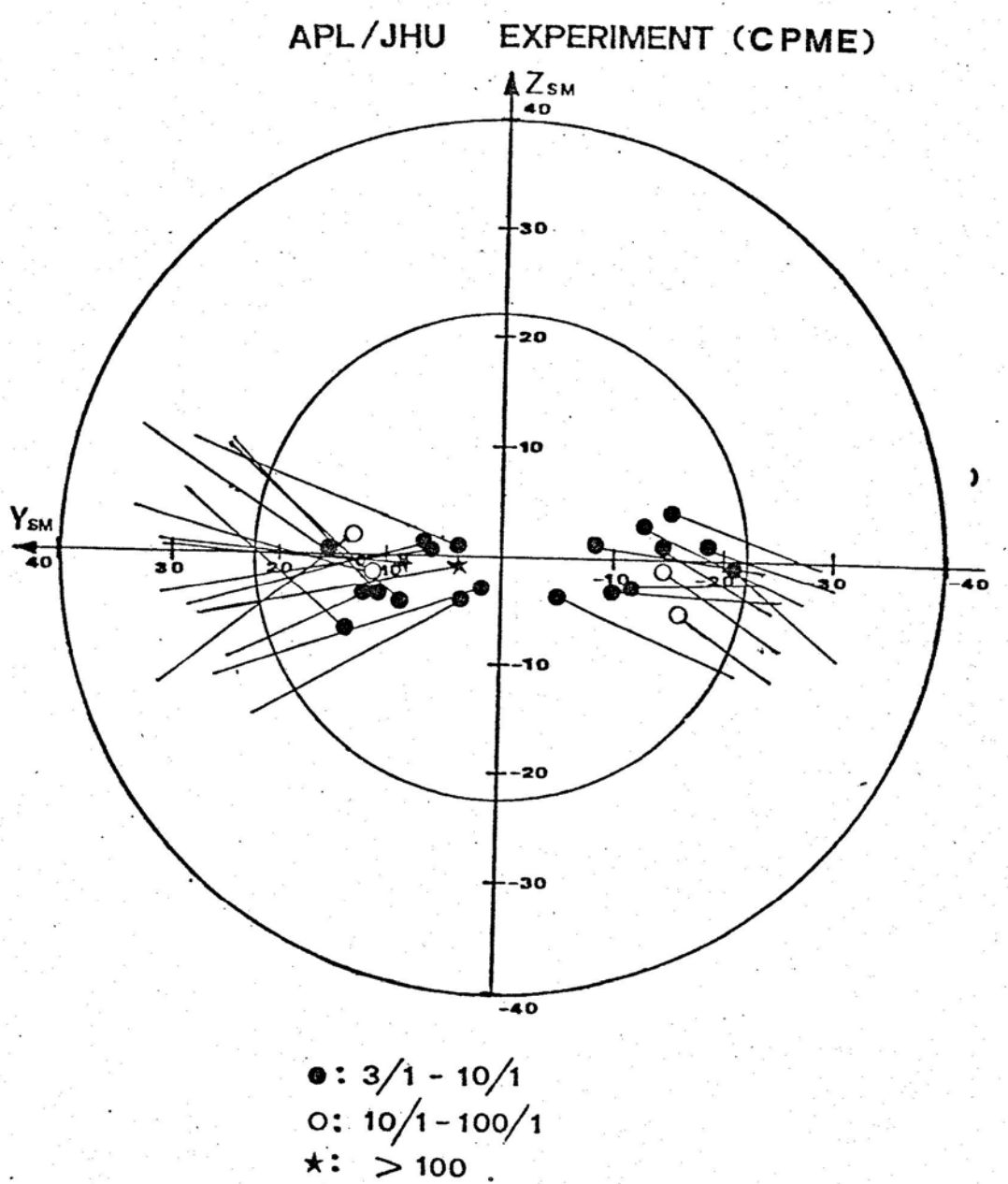

**Fig.7** Presents the comparison of magnetic particle burst observed by two spacecrafts, one been inside the plasma sheet and the other been out of the magnetosphere and the magnetosheath. These observations clearly indicates strong active regions inside the plasma sheet. The active regions in plasma sheet correspond to non-equilibrium fractal dissipation of magnetic energy and fractal acceleration of energetic particles.